\documentclass[preprint]{aastex}
\newcommand{\lam}{$\lambda$}

\renewcommand{\ion}[2]{#1\,{\sc #2}}
\newcommand{\ecs}{erg\,cm$^{-2}$\,s$^{-1}$ sr$^{-1}$} % erg/cm2/s
\newcommand{\as}{${^\prime}{^\prime}$}

\shorttitle{New Fe\,IX line identifications}
\begin{document}


\title{New EUV Fe\,IX emission line identifications from Hinode/EIS}


\author{P. R. Young}

% \altaffiltext{1}{Space Science Division, Naval Research Laboratory,
%   Washington, DC 20375}
% \altaffiltext{2}{George Mason University, 4400 University Drive, Fairfax, VA 22030}
% \altaffiltext{3}{Artep, Inc. Ellicott City, MD 21042}

\affil{Space Science Division, Naval Research Laboratory,
  Washington, DC 20375}
\affil{George Mason University, 4400 University Drive, Fairfax, VA 22030}


\begin{abstract}
Four \ion{Fe}{ix} transitions in the wavelength range 188--198~\AA\
are identified for the first time in spectra from the EUV Imaging
Spectrometer on board the Hinode satellite. In particular the emission
line at 197.86~\AA\ is unblended and close to the peak of the EIS
sensitivity curve, making it a valuable diagnostic of plasma at around
800,000~K -- a critical temperature for studying the interface
between the corona and transition region. Theoretical ratios amongst the four
lines predicted from the CHIANTI database reveal weak sensitivity to
density and temperature with observed values consistent with
theory. The ratio of  
\lam197.86 relative to the  \lam171.07 resonance line of \ion{Fe}{ix} 
is found to be an excellent temperature diagnostic, independent
of density, and the derived temperature in the analysed data set is
$\log\,T=5.95$, close to the predicted temperature of maximum
ionization of \ion{Fe}{ix}. 
\end{abstract}


\section{Introduction}

Although the solar vacuum ultraviolet (VUV; 100--2000~\AA) spectrum
has been studied for many years and most of the strong lines identified,
there remain a large number of unidentified lines. Classifying these
lines is important for several reasons: they may yield new diagnostics
for the Sun or other astrophysical bodies; they may contribute to the
passbands of solar imaging instruments, distorting results if not
accounted for;
they add to the total irradiance from the
Sun; and they improve our knowledge of atomic structure.
While line identifications can be performed in the laboratory, a high
resolution solar spectrometer with imaging capability is a 
particularly powerful tool for 
line identification studies as the morphology of images formed in a
given emission line will be highly characteristic of the temperature
the line is formed at, making the parent ion much easier to establish.

Recently the EUV Imaging
Spectrometer \citep[EIS;][]{culhane07} was launched on board the
Hinode satellite \citep{kosugi07} and it is the first space-borne
spectrometer to routinely observe the Sun in the wavelength ranges
170--212 and 246--292~\AA\ at a high spectral resolution. The majority
of strong emission lines and their diagnostic properties are known
\citep{young07b, brown08}, but 45~\% of lines were classified as
unidentified by \citet{brown08}. The majority of these unidentified
lines likely belong to the iron ions \ion{Fe}{ix--xiv} for which there
are many levels that do not have experimental energy levels, thus
making the wavelengths from these emission lines very uncertain.

This work presents new identifications for four such emission lines
that belong to \ion{Fe}{ix}. Although the strong \lam171.07 resonance
line of \ion{Fe}{ix} is found in the EIS short wavelength band, the
effective area is so low (around 1000 times less than the peak
effective area of the instrument) that it is not scientifically useful
in most circumstances. The newly-identified lines are much stronger in
terms of counts at the detector and bring new capability to the
instrument, giving access to strong unblended lines from the complete
sequence of iron ions from \ion{Fe}{viii} to \ion{Fe}{xvii}. In
addition, the newly-identified line at 197.86~\AA\ is found to
contribute to
one of the best coronal temperature diagnostics at ultraviolet
wavelengths.

\section{Observations}

\begin{figure}
\epsscale{0.90}
\plotone{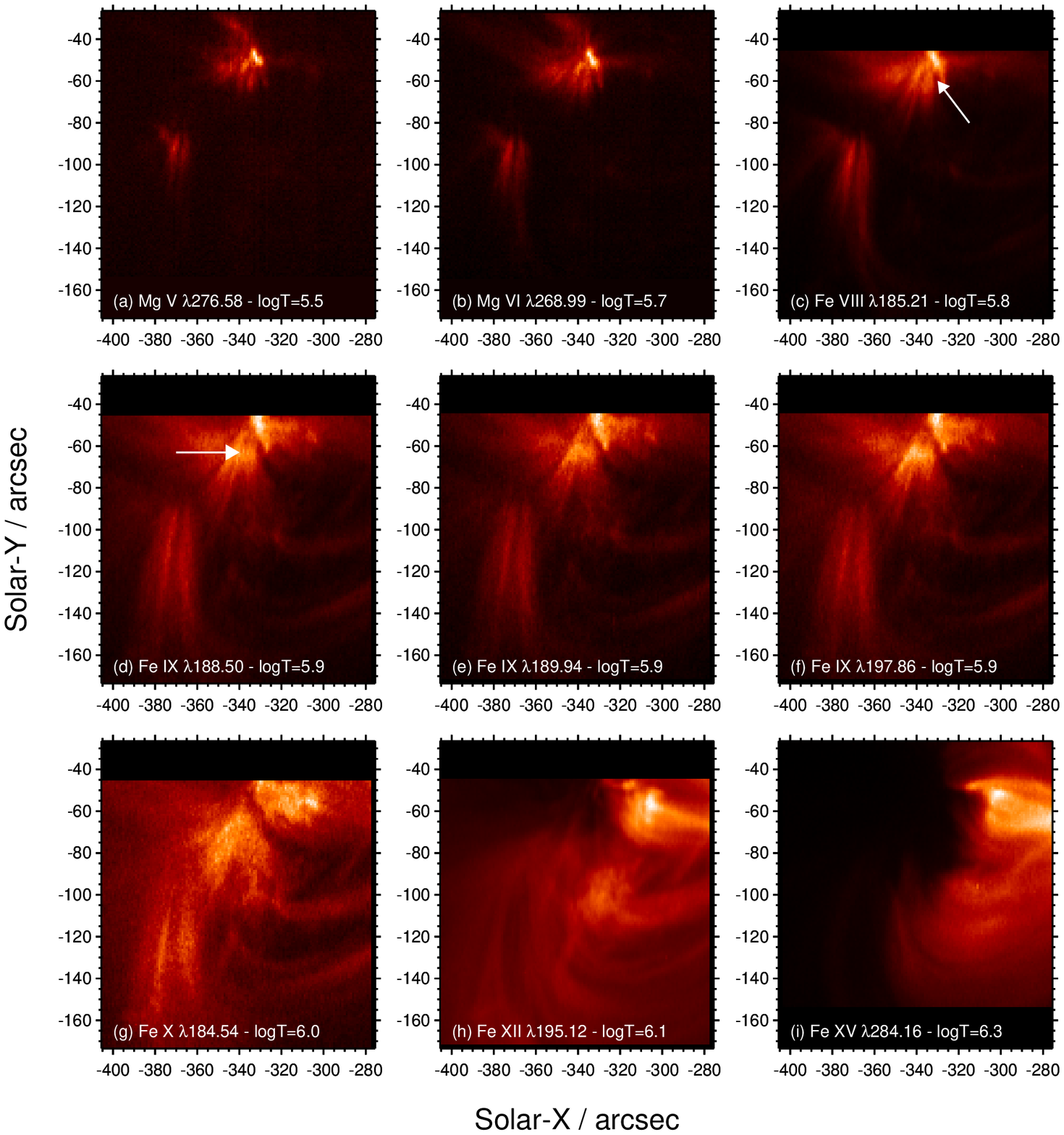}
\caption{Hinode/EIS images in nine emission lines from the 2007
  February 21 data set. Panels (d)--(f) show the images from the
  newly-identified \ion{Fe}{ix} lines. The images have been corrected
  for the various spatial offsets that are a function of wavelength,
  hence the black bars in the images. The significance of the arrows
  is explained in the main text.\label{fig.images}} 
\end{figure}

Active region AR 10942 was observed by EIS on 2007 February 21 with
the study HPW008\_FULLCCD\_RAST which covers an area 128\as\ $\times$
128\as\ with the 1\as\ slit. Full CCD spectra are obtained with an
exposure time of 25~s, and the raster duration is 57~mins. Images from
the data-set in a variety of emission lines are shown in
Fig.~\ref{fig.images}. The raster is pointed at the footpoint regions
on the east side of the active region.
A number of spiky structures are seen in the
cool lines which are seen to be the footpoints of loops visible in
\ion{Fe}{x} and \ion{Fe}{xii}.
A key
feature of this observation is a spatial area (indicated by an arrow
in Fig.~\ref{fig.images}c) that is bright at
temperatures $\log\,T=5.7$--6.0, but has relatively little emission
from plasma at $\log\,T>6.0$. The spectrum in this region is not dissimilar to a
coronal hole spectrum, but because the observed features are bright coronal
loop footpoints the signal-to-noise is high in the spectrum allowing
weak lines to be observed.

The data set was calibrated using the routine EIS\_PREP, which is
described in detail by \citet{young08}. In order to measure the
intensities and wavelengths of the \ion{Fe}{ix} lines, a region
of 38 spatial pixels indicated by the arrow in Fig.~\ref{fig.images}d
was chosen where the \ion{Fe}{ix} lines are 
particularly enhanced. The spectra from these pixels were averaged,
with care taken not to include pixels flagged as `missing' by
EIS\_PREP. The spatial offsets between different wavelengths
highlighted by \citet{young08} that are due to the tilt of the EIS
grating relative to the EIS CCD have been corrected for to ensure the
same spatial region is used for each line.

The present work is focussed on four emission lines at 188.50, 189.94,
191.22 and 197.86~\AA\ that were seen to be strongly enhanced in the
loop footpoint regions. Only one of these (\lam197.86) was classified as
unidentified by \citet{brown08}, with 
\lam188.50 listed as a \ion{Fe}{xii} transition, \lam189.94 as a \ion{Fe}{xi}
transition and \lam191.22 as a blend of \ion{Fe}{xiii} and
\ion{S}{xi}. 
To identify the lines with \ion{Fe}{ix} we compare images formed in
the lines with other, known species.
Fig.~\ref{fig.images} shows images from \ion{Mg}{v} \lam276.58,
\ion{Mg}{vi} \lam268.99, \ion{Fe}{viii} \lam185.21,
\ion{Fe}{x} \lam184.54, \ion{Fe}{xii} 
\lam195.12 and \ion{Fe}{xv} \lam284.16, together with images from the
lines at 188.50, 189.94 and 197.86~\AA.  (The 191.22~\AA\ image looks
very similar, except there is contribution to the image from hotter
plasma. This is due to the blending \ion{S}{xi} line -- see below.)
In each case the
images were formed by summing 7 wavelength pixels across the line
profile.

Comparing the images it is clear, firstly, that \lam188.50, \lam189.94
and \lam197.86 are
formed at the same temperature since the images are very similar to
each other and, secondly, that the  temperature of formation is
somewhere between that of \ion{Fe}{viii} and
that of \ion{Fe}{x}. As the unidentified lines are strong (within a factor 2
in intensity of the strongest \ion{Fe}{viii} lines observed by EIS)
they must belong to an ion with a high solar abundance. Since line
identifications for simple ions are well known at EUV wavelengths, the
most likely ion is \ion{Fe}{ix}.  The fact that the \lam188.50 and
\lam189.94 images look almost identical to \lam197.86 (an apparently
unblended line) suggests that \ion{Fe}{xii} and \ion{Fe}{xi} make very
little contribution to \lam188.50 and \lam189.94,
respectively. \ion{S}{xi} \lam191.26 is partly blended with \lam191.22
and in fact dominates the \ion{Fe}{ix} line in the cores of active
regions. Note that the \ion{Fe}{ix} component compromises the use of
\ion{S}{xi} \lam191.26 in de-blending the \ion{Fe}{xii} \lam186.88
density diagnostic line \citep{young07b}. 
The CHIANTI database
\citep{landi06,dere97} predicts that the \ion{Fe}{xiii} line at
191.2~\AA\ will be negligible
in all conditions.

The measured wavelengths and line intensities of the four \ion{Fe}{ix}
transitions  are given in Table~\ref{tbl.transitions},
together with the
newly-suggested line identifications. The details for the measured
\lam171.07 line are also given. 
The rest wavelengths of the \ion{Fe}{ix} lines are derived by first
measuring the wavelengths of \ion{Fe}{x} \lam184.54 and \lam190.04 in
the spectrum. The velocity shifts of both lines relative to the rest
wavelengths given in \citet{brown08} are then derived and
averaged. The measured wavelengths of the \ion{Fe}{ix} lines are then
corrected by subtracting the \ion{Fe}{x} velocity shift. This process
assumes that \ion{Fe}{ix} and \ion{Fe}{x} exhibit the same velocity in
the data set which we feel is reasonable given their close
temperatures of formation.

\begin{table*}[h]
\begin{center}
\caption{\ion{Fe}{ix} transitions observed by EIS.\label{tbl.transitions}}
\begin{tabular}{lllll}
\tableline\tableline
Rest&&CHIANTI & Observed & Predicted \\
Wavelength & Transition & wavelength & Intensity & Intensity\tablenotemark{a} \\
(\AA) &&(\AA) &(\ecs) &(\ecs) \\
\tableline
171.071 & $3s^23p^6$ $^1S_0$ -- $3s^23p^53d$ $^1P_1$ &
     171.073 &5794 & 7328 \\
188.497 & $3s^23p^53d$ $^3F_4$ -- $3s^23p^4(^3P)3d^2$ $^3G_5$ &
     174.245 & 341 & 341 \\
189.941 & $3s^23p^53d$ $^3F_3$ -- $3s^23p^4(^3P)3d^2$ $^3G_4$ &
     175.477 & 201 & 206 \\
191.216 & $3s^23p^53d$ $^3F_2$ -- $3s^23p^4(^3P)3d^2$ $^3G_3$ &
     176.660  & 92 & 85\\
197.862 & $3s^23p^53d$ $^1P_1$ -- $3s^23p^54p$ $^1S_0$ & 
     173.149 & 192 & 219 \\
\tableline
\end{tabular}
\tablenotetext{a}{Predicted intensities are from the CHIANTI synthetic
  spectrum discussed in the text, normalized relative to \lam188.497.}
\end{center}
\end{table*}

\begin{figure}
\plotone{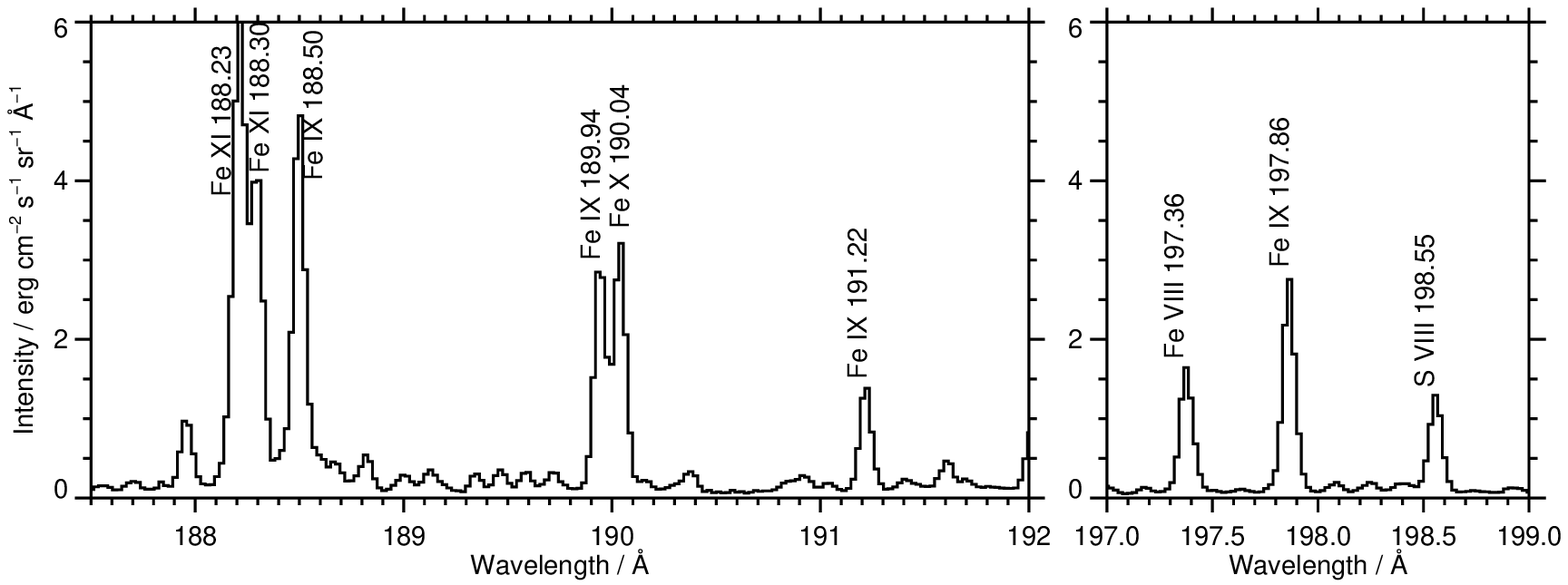}
\caption{EIS spectra from the 2007 February 21 data set showing the
  newly-identified \ion{Fe}{ix} lines.\label{fig.spectrum}}
\end{figure}

\section{Atomic data and emission line modeling}

The CHIANTI atomic database  contains models
suitable for predicting emission line spectra from
most astrophysically abundant ions. The \ion{Fe}{ix} model consists of
observed energies from v.3.0 of the NIST Atomic Database, and theoretical
energies, Maxwellian-averaged electron collision strengths and
radiative decay rates from \citet{storey02}. The model has 140 levels
from six configurations.

To help identify the \ion{Fe}{ix} transitions a synthetic isothermal spectrum was
generated using CHIANTI at a temperature
$\log\,T=5.9$ and electron density $\log\,N_{\rm
  e}=9.0$. Although the ionization balance
calculations of \citet{bryans06} give a temperature of maximum
ionization, $T_{\rm max}$, of $\log\,T=5.8$ for \ion{Fe}{ix}, we
believe it is actually $\log\,T=5.9$: from EIS observations \ion{Fe}{viii} clearly has
$\log\,T_{\rm max}=5.8$ \citep{young07a} yet \citet{bryans06} give
$\log\,T_{\rm max}=5.6$. \ion{Fe}{ix} must thus be principally formed
at $\log\,T=5.9$. From the EIS spectrum obtained in the \ion{Fe}{ix}
emitting region, we used the \ion{Mg}{vii} \lam280.75/\lam278.39
density diagnostic \citep{young07a}, formed at a slightly cooler
temperature of $\log\,T_{\rm max}=5.8$ compared to \ion{Fe}{ix}, to derive a density of $\log\,N_{\rm
  e}=9.0$, and so this value was used.
The synthetic spectrum reveals a number of lines clustered around the \lam171.07
resonance line between 155 and 185~\AA, several of which have
intensities comparable to the $3s^23p^6$--$3s^23p^53d$ transitions
\lam217.10 and \lam244.91, and thus would be expected to be observed if they
fall in the EIS wavelength range. Every one of the emission lines
between 155 and 185~\AA\ (except \lam171.07) only
have theoretical wavelengths, and all but two of them are
$3s^23p^53d$--$3s^23p^43d^2$ transitions. 

As found in the previous section, the four unidentified EIS lines have
formation temperatures close to that of \ion{Fe}{ix} thus we look to
identify these with the predicted CHIANTI lines. The strongest of the
lines between 155 and 185~\AA\ we identify with the strongest of the
EIS lines, namely \lam188.50. As this is a transition between levels
$^3G_5$ and $^3F_4$ then, based on standard atomic physics properties
for multiplets,
we expect weaker $^3F_3$--$^3G_4$ and $^3F_2$--$^3G_3$ transitions whose
wavelength separations will be fairly similar to the separations in
CHIANTI. The wavelengths and intensities given in
Table~\ref{tbl.transitions} demonstrate that \lam189.94 and \lam191.22
are an excellent match for these transitions.

The closest CHIANTI transition to match the intensity of the feature
at 197.86~\AA\ is $3s^23p^53d$ $^1P_1$ -- $3s^23p^54p$
$^1S_0$, whose theoretical wavelength is 173.149~\AA. As this is a
singlet--singlet transition then there are no companion lines to serve
as a check, but the diagnostic ratios presented in the following
section are consistent with the identification.

New energy values for the upper levels of the new identifications are
provided in Table~\ref{tbl.energies}. We note that the differences
with the theoretical energies of \citet{storey02} are 4~\%\ and 6~\%\
for the $3s^23p^43d^2$ $^3G$ and $3s^23p^54p$ $^1S$, respectively, which are
consistent with the accuracy expected of theoretical energy
calculations at these energies.
The error bars on the energy estimates are obtained as
follows: for $3s^23p^54p$ $^1S_0$ the energy is derived from the
wavelength of \lam171.07 and the wavelength of \lam197.86. The latter
is derived here with an accuracy of $\pm$0.005~\AA, while the rest
wavelength of the former
is taken from \citet{behring76} who give an accuracy of $\pm$0.004~\AA.
For the $3s^23p^53d$ $^3F_J$ -- $3s^23p^4(^3P)3d^2$ $^3G_{J^\prime}$
transitions, the energies of the lower levels are provided by
\citet{edlen78}. Since none of the $^3F_J$ levels decay to the ground
level, their energies are derived indirectly through allowed transitions
to the $3s3p^63d$ configuration, and allowed transitions from this
configuration to the ground $3s^23p^6$ configuration. Uncertainties are not provided by
\citet{edlen78} but we can estimate values of $\pm$15~cm$^{-1}$ from
the wavelength errors provided by \citet{smitt83}. Combining
with the uncertainties in the measured EIS wavelengths of
$\pm$0.005~\AA\ yields the uncertainties in Table~\ref{tbl.energies}.

% NOTE: predicted wavelengths: 3F3-3G3 is $189.582\pm 0.011$; 3F4-3G4 is
% $188.686\pm 0.011$.

\begin{table}[h]
\begin{center}
\caption{New \ion{Fe}{ix} energy levels.\label{tbl.energies}}
\begin{tabular}{lll}
\tableline\tableline
Level & Energy / cm$^{-1}$  & Index\tablenotemark{a} \\
\tableline
$3s^23p^4(^3P)3d^2$ $^3G_4$ & $955\,790\pm 20$ &94 \\
$3s^23p^4(^3P)3d^2$ $^3G_5$ & $956\,322\pm 21$ &95 \\
$3s^23p^4(^3P)3d^2$ $^3G_3$ & $956\,788\pm 20$ &96 \\
$3s^23p^54p$ $^1S_0$ & $1\,089\,949\pm 19$   &140 \\
\tableline
\end{tabular}
\tablenotetext{a}{The level index in the CHIANTI v5.2 model.}
\end{center}
\end{table}

\section{Diagnostic possibilities}

\begin{figure}
\plotone{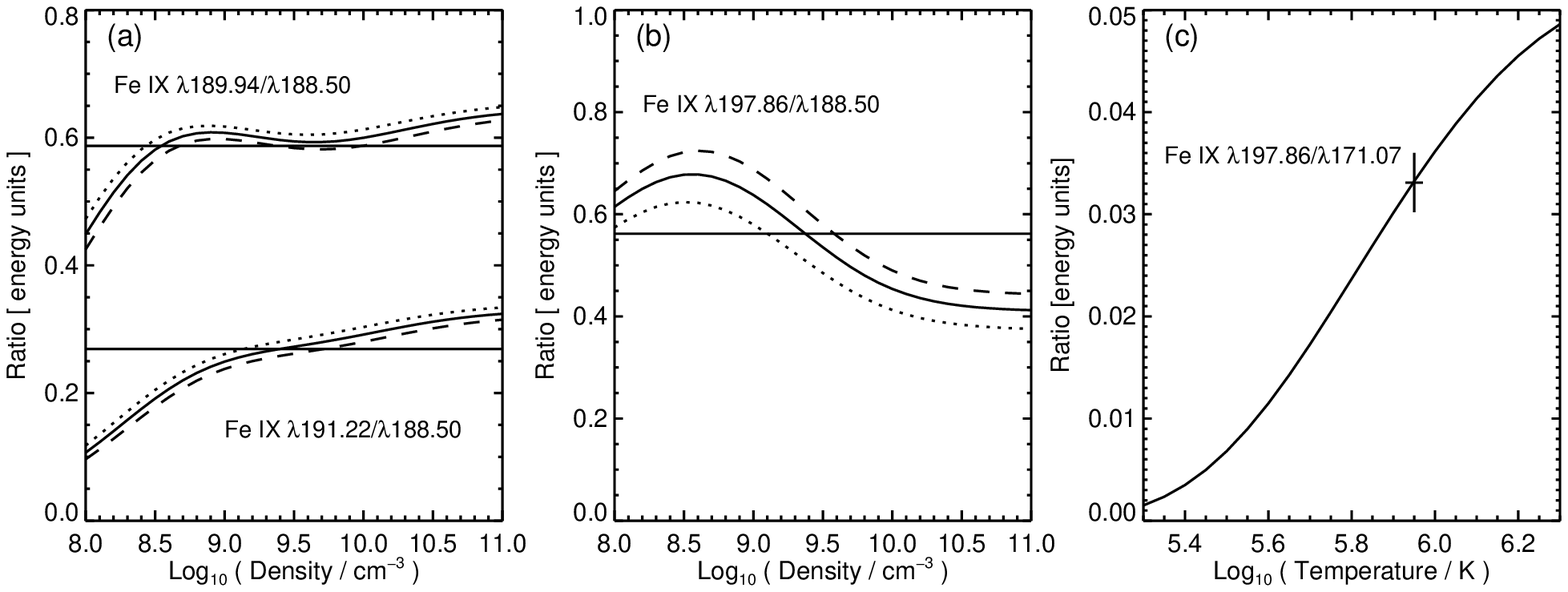}
\caption{Panels (a) and (b) show predicted line ratio variations from
  CHIANTI that involve the four
  newly-identified \ion{Fe}{ix} transitions, relative to
  \lam188.50. Ratios for three different temperatures --
  $\log\,T=5.8$ (dotted line), 5.9 (solid line) and 6.0 (dashed line)
  -- are shown. The horizontal lines show the measured values of each
  ratio. Panel (c) shows the varation of the \lam197.86/\lam171.07
  ratio with temperature, calculated at $\log\,N_{\rm e}=9.0$. The
  measured ratio value with error bars is indicated.\label{fig.diags}}
\end{figure}

Amongst the newly identified \ion{Fe}{ix} lines, the CHIANTI model
predicts some density and temperature
sensitivity. Figs.~\ref{fig.diags}(a,b) show the
\lam191.22/\lam188.50, \lam189.94/\lam188.50 and \lam197.86/\lam188.50
ratios as a function of density for temperatures $\log\,T=5.8$, 5.9
and 6.0. The horizontal lines in the plots indicate the measured
ratios from the 2007 February 21 data set and good agreement is found
with the density derived from the slightly cooler \ion{Mg}{vii}
density diagnostic.

In Fig.~\ref{fig.diags}(c) the CHIANTI prediction for the
\lam197.86/\lam171.07 ratio as a function of temperature is given. The
ratio is an excellent temperature diagnostic and we find a temperature
of $\log\,T=5.95\pm 0.05$ from the spectrum analysed here. This is
close to the $\log\,T_{\rm max}$ value of 5.9 discussed earlier,
giving confidence in our identification of the \lam197.86 line.
This \ion{Fe}{ix} diagnostic is of great
interest as the lines are close in wavelength and have very little
density sensitivity, making it one of the best temperature diagnostics
available in the VUV.

\section{Summary}

Four new \ion{Fe}{ix} line identifications in the wavelength range 188
to 198~\AA\ have been presented, based on spectra obtained by the
Hinode/EIS instrument. The lines give temperature coverage for the EIS
instrument at $\log\,T=5.9$, a region not covered by any other
ion species observed by EIS, and one of great interest as it is where
the transition region gives way to the corona.

The identified \lam188.50 and \lam197.86 lines are comparable
in strength in terms of counts measured on the detector, but
\lam197.86 is preferred for observations as it appears to be unblended
and there are no strong lines nearby to it. Since \lam188.50 lies
close to the strong \ion{Fe}{xi} \lam\lam188.23, 188.30 lines then many observing
studies already obtained with EIS will have observed this
line. E.g., a wavelength window set to 32 pixels and centered on
188.23~\AA\ will include wavelengths up to 188.60~\AA.

\lam197.86/\lam171.07 is found to be an excellent temperature
diagnostic, with very little density sensitivity and lines relatively
close in wavelength. The ratio is difficult to use from EIS data  as
the EIS sensitivity is very low at 171~\AA, but designs for a future
EUV solar spectrometer should give consideration to observing both lines.

\acknowledgments

E.~Landi is thanked for useful comments on the manuscript.
Hinode is a Japanese mission developed and launched by
ISAS/JAXA, with NAOJ as domestic partner and NASA and
STFC (UK) as international partners. It is operated by
these agencies in co-operation with ESA and NSC (Norway).


%% in parentheses, after the keyword, but they will not be verified.

{\it Facilities:} \facility{Hinode (EIS)}.



\end{document}